\documentclass[%
 aip,
 cha,
 amsmath,amssymb,
 twocolumn, 
 reprint,%
]{revtex4-1}

\usepackage{graphicx}
\usepackage{dcolumn}
\usepackage{bm}
\usepackage[utf8]{inputenc}
\usepackage[T1]{fontenc}
\usepackage{mathptmx}
\usepackage{etoolbox}
\usepackage{lipsum}
\usepackage[usenames,dvipsnames]{color}
\usepackage[colorlinks=true,linkcolor=blue,citecolor=blue,urlcolor=blue]{hyperref}
\usepackage{tikz}
\usepackage{cleveref}
\usepackage{booktabs} 
\usepackage{longtable} 
\usepackage{xcolor}
\definecolor{darkgreen}{rgb}{0.0, 0.5, 0.0}

\makeatletter
\def\@email#1#2{%
 \endgroup
 \patchcmd{\titleblock@produce}
  {\frontmatter@RRAPformat}
  {\frontmatter@RRAPformat{\produce@RRAP{*#1\href{mailto:#2}{#2}}}\frontmatter@RRAPformat}
  {}{}
}%
\makeatother

\begin{document}

\preprint{AIP/123-QED}

\title{\texorpdfstring{Hierarchical Clustering in Mean-Field Coupled Stuart-Landau Oscillators}{Hierarchical Clustering in Mean-Field Coupled Stuart-Landau Oscillators}}
\author{Nicolas Thomé}
\email{nicolas.thome@tum.de}
\affiliation{School of Natural Sciences, Nonequilibrium Chemical Physics, Technische Universität München, James-Franck-Str. 1, D-85748 Garching, Germany}
\author{Matthias Wolfrum}
\affiliation{Weierstrass Institute for Applied Analysis and Stochastics, Mohrenstrasse 39, DE-10117 Berlin, Germany}
\author{Katharina Krischer}
\email{krischer@tum.de}
\affiliation{School of Natural Sciences, Nonequilibrium Chemical Physics, Technische Universität München, James-Franck-Str. 1, D-85748 Garching, Germany}

\date{\today}

\begin{abstract}
\textbf{Clustered solutions in oscillator networks provide an important insight into how a system might diversify from a synchronous solution into spatiotemporal complex solutions. They can therefore form a link between fully synchronized and incoherent states. Despite their fundamental role in coupled oscillator dynamics, our understanding of how these clusters form and differentiate is still quite limited. Here, we study an ensemble of globally coupled Stuart-Landau oscillators and focus on the question of how 3-cluster solutions emerge from 2-cluster solutions and how the different 3-cluster solutions are organized in parameter space. We show that the arrangement of the clusters is dictated by a co-dimension 2 point, which we coin Type-II cluster singularity. Furthermore, our study points to a hierarchical structure of higher cluster solutions.} 

\end{abstract}

\maketitle
\begin{quotation}
Since the seminal work of A. Turing\cite{Turing.1990}, the self-organized formation of patterns in homogeneous media has become a prominent topic in nonlinear science. Classical results refer to chemical reactions in continuous diffusive media. Much less is known about the pattern formation for oscillatory processes and global interaction via a mean field. In this case, spontaneous emergence of cluster structures is an important phenomenon. A paradigmatic system for such processes are globally coupled Stuart-Landau oscillators. As in classical Ginzburg-Landau theory, the interplay of the shear parameters in the nonlinearity and in the coupling can induce a multitude of dynamical phenomena, which display a transition from fully synchronous to completely incoherent behavior. Within this transition one can observe how gradually both the spatial and the temporal complexity increases. 
\end{quotation}
\section{\label{sec:Intro}Introduction}
The study of oscillator-network systems is crucial for understanding a plethora of natural phenomena, from collective dynamics in neural networks \cite{Bick.2020} to the coordinated beating of cardiac cells\cite{Winfree.1967} and the functioning of power grids\cite{Albert.2004}. By investigating the intricate dynamics of coupled oscillators, we can gain insights into the fundamental principles that govern collective behavior in complex systems. \\
Interacting networks of oscillators exhibit a rich diversity of dynamical behaviors, ranging from synchronous solutions, where all oscillators move in unison, to completely incoherent solutions, where no discernible synchronization pattern exists\cite{Rosenblum.2003, Strogatz.2000}. Intriguing intermediate states 
lie between these extremes.
They are characterized by the self-organized formation of fully synchronized subpopulations, called {\em clusters}, which show increasingly complex interpopulation dynamics and possibly also individual oscillators 
that move on their own trajectories and do not belong to any cluster. 
The hierarchical structure among these clusters is particularly significant as it organizes the transition from synchrony to incoherence\cite{Haugland.2021, Iwasa.2010}. An essential step is understanding how a coupled oscillator system bifurcates from synchronous oscillations into the differentiated dynamics of cluster solutions\cite{Schmidt.2015}. \\
Various networks of oscillators have been investigated which differ in the coupling topology, e.g., non-local coupling\cite{Kuramoto.1984} or randomized network architecture\cite{Um.2014}, or the type of oscillators, e.g., phase oscillators\cite{Kuramoto.2002} or relaxation oscillators\cite{Christiansen.1993}. Here, we work with globally, i.e. all-to-all, coupled identical Stuart-Landau oscillators. The Stuart-Landau (SL) oscillator is the prototypical 2D oscillator, where the amplitude-phase interaction allows for a particularly rich cluster dynamics. Global coupling provides a simplified yet robust framework and has been subject of many previous studies on ensembles of SL oscillators, allowing us to build on a comparatively large body of results \cite{Hakim.1992, Chabanol.1997, Nakagawa.1993, Nakagawa.1994, Kemeth.2018, Kemeth.2019, Ku.2015, Daido.2007}. \\
The governing equation of a system of SL oscillators under linear global coupling reads: 
\begin{equation}
\label{eq:GCSLO}
\begin{aligned}
    \partial_t W_k &= W_k - (1+iC_2) \big|W_k \big|^2W_k \\
    &\quad + K(1+iC_1)\left(\frac{1}{N}\sum_{l=1}^{N}W_l - W_k\right),
\end{aligned}
\end{equation}
where $W_k(t)\in \mathbb{C}$ is a complex, time-dependent variable representing an oscillator indexed by $k \in \{1,..,N\}$. $K, C_1, C_2\in \mathbb{R}$ are free  parameters. $C_2$ determines the eigenfrequency of the identical oscillators, $K$ and $C_1$ define together the complex-valued global coupling strength\cite{Hakim.1992}. 
\newline
Analytical solutions were obtained for situations in which the individual oscillators decouple. This occurs in two ways: (a) For strong, attractive coupling, all oscillators behave identically such that the coupling term vanishes and 
all the oscillators rotate together along the unit circle. (b) If the coupling is not strong enough, $\frac{1}{N}\sum_{l=1}^{N}W_l$ may vanish and all oscillators are frequency-locked and oscillate with an incoherent phase distribution and an amplitude that is in general different from one. The stability of the synchronous and the vanishing-mean solution was determined using linear stability analysis \cite{Nakagawa.1993, Chabanol.1997}. The stability boundary of the synchronous solution is typically called the Benjamin-Feir instability. 
\begin{figure}
\includegraphics{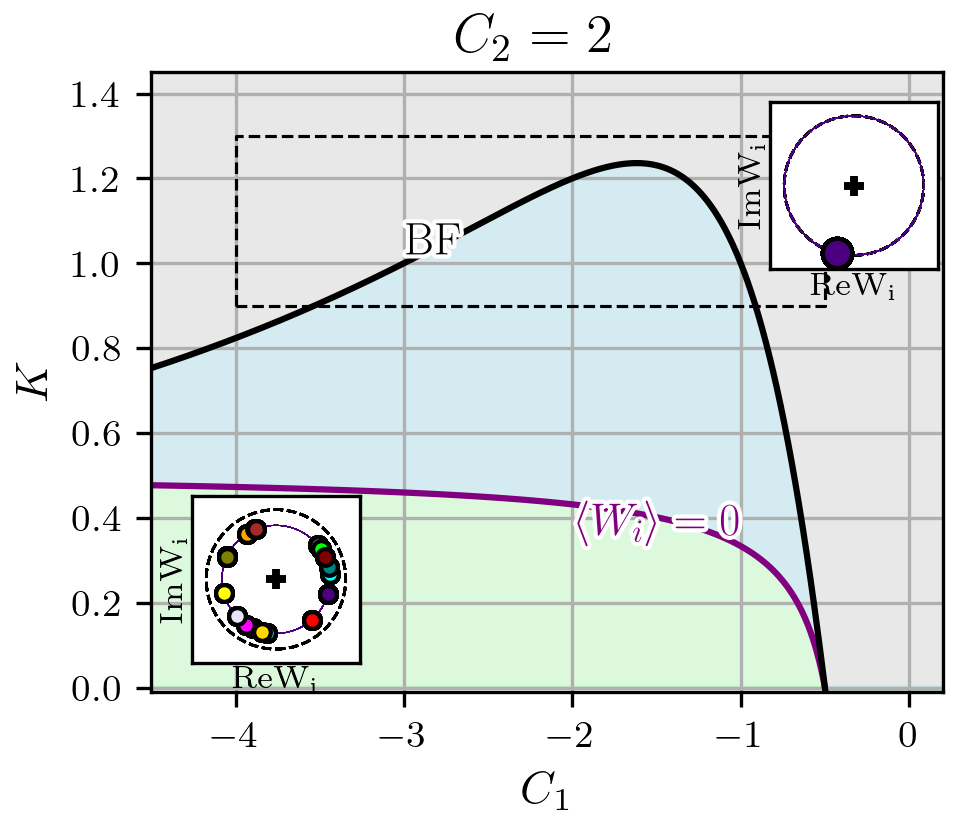}
\caption{\label{fig:PrevRe}Bifurcation diagram of a system of globally coupled Stuart-Landau oscillators with linear global coupling for $C_2=2$. The black line depicts the Benjamin-Feir (BF) instability, at which the synchronous solution loses stability. The synchronous solution is stable in the gray area. The purple line indicates the stability boundary of the vanishing-mean solution, which is stable in the green region. The two insets show snapshots of typical synchronous and vanishing-mean solutions in the complex plane. The dashed circle indicates the unit circle. In the blue area, a multitude of more complex solutions exist. The dashed rectangle shows the parameter region that is depicted in Figure \ref{fig:Heatmap10}.}
\end{figure}
In Figure \ref{fig:PrevRe}, we show the stability regions of the synchronous solution (gray area) and the vanishing mean solution (green area) in the parameter plane spanned by the two coupling constants $K$ and $C_1$ for $C_2=2$, following ref. \cite{Nakagawa.1993}. The insets in the upper right and the lower left corner depict typical snapshots of these two solutions in the complex plane, where the dashed circles indicate the unit circle. The black line in Figure \ref{fig:PrevRe} represents the Benjamin-Feir instability, the purple line the stability boundary of the vanishing mean solution. In the blue region between these two stability boundaries, neither the synchronous nor the vanishing mean solution is stable, but a "rich variety of collective behaviors"\cite{Nakagawa.1994} exists. Various studies have described different synchronization patterns in this parameter range, such as 2- and 3-cluster configurations, collective chaos, chimera states, or oscillator death \cite{Nakagawa.1993, Nakagawa.1994, Sethia.2014, Bi.2014}.
However, much less is known about how these states arise and how they are related. \\
Best studied are 2-cluster solutions. Ku et al.\cite{Ku.2015} investigated the transition between different 2-cluster solutions and showed that they are hysteretic. Using symmetry arguments, Banaji\cite{Banaji.2002} characterized the instability of the synchronized solution as a highly degenerate symmetry breaking bifurcation in which 
2-cluster states with all different cluster sizes emerge simultaneously on a multitude of coexisting transcritical or pitchforking branches.
 Furthermore, the authors showed that the 2-cluster solutions are born in saddle-node bifurcations. Based on this work, Kemeth et al.\cite{Kemeth.2019} demonstrated that 2-cluster solutions are organized in a co-dimension-2 bifurcation, which the authors dubbed "cluster singularity". At the cluster singularity, cascades of bifurcations emerge, most notably a cascade of saddle-node bifurcations creating 2-cluster solutions of all possible cluster sizes, 
 and two cascades of transverse bifurcations that stabilize and destabilize these states, respectively. Employing a center manifold reduction combined with a normal-form approach helped to further elucidate the properties and intricacy of this codimension-2 point \cite{Kemeth.2020}. 
Further away from the cluster singularity in parameter space, the dynamics of the 2-cluster states are still poorly understood. To explore the transition from symmetric to less symmetric solutions in 2-cluster formations, the bifurcation pathways and dynamics must be analyzed in greater depth, particularly in the context of emergent structures and turbulence. In essence, the path from the 2-cluster solutions resulting from cluster singularity to incoherence must be uncovered. Ref. \cite{Kemeth.2018} gives a preliminary look at bifurcation scenarios for $N = 4$ without putting it into the context of the cluster singularity.\\
Our work aims to fill this gap by investigating how 2-cluster solutions “specify” into distinguishable subclusters when symmetry breaks down \cite{Golubitsky.2002}; how once indistinguishable clusters diverge into different formations and evolve into 3-cluster solutions, as a first step towards turbulent states.
\section{\label{sec:Prelimiaries} Preliminaries}
In this section we recall basic concepts and techniques for systems with symmetry and describe the theoretical background for our treatment of the cluster dynamics of system \eqref{eq:GCSLO}.
\paragraph{Symmetries, invariant subspaces, and reduced systems }
System (\ref{eq:GCSLO}) has two types of symmetries. First, since all the oscillators are identical, it is invariant under permutations of the indices. This implies that for any solution swapping of two oscillators will produce again a solution. In other words, system (\ref{eq:GCSLO}) is equivariant under the symmetry group $\mathbf{S}_N$ of all permutations of the indices $k=1\dots N$. Second, the system is equivariant under phase shifts, such that when $\vec W(t)$ is a solution of (\ref{eq:GCSLO}), then for all $\chi\in S^1=\mathbb{R}/2\pi\mathbb{Z}$ also $\vec W(t)\exp(i\chi)$ is a solution of Eqs. (\ref{eq:GCSLO})\cite{Hoyle.2006}.

The cluster solutions are a consequence of the permutation symmetry of the system. This can be seen as follows. Whenever two oscillators have an identical initial condition, i.e., $W_k(0)=W_j(0)$ for some $1\leq k<j\leq N$, then the whole trajectory has this property and $W_k(t)=W_j(t)$ for all $t\in\mathbb{R}$. 
In this way, the permutation symmetry leads to a hierarchy of invariant subspaces
$$W_1=\dots =W_{N_1},W_{N_1+1}=\dots =W_{N_1+N_2},W_{N_1+N_2+1}=\cdots,$$
called {\em cluster subspaces}, where $N_1\dots N_n$ are the cluster sizes with $\sum_j N_j=N$, and $n$ is the number of clusters. A solution in such a cluster subspace is called an {\em $n$-cluster solution} and an oscillator in a trivial cluster with $N_j=1$ is called a {\em solitary oscillator}\cite{Schulen.2022,Berner.2020,Jaros.2018}. Obviously, the situation with $n=1$ corresponds to full synchrony. 

The dynamics within such a cluster subspace is given by
\begin{align}\label{eq:red}
    \partial_t \hat W_j &= \hat W_j - (1+iC_2) \big|\hat W_j \big|^2\hat W_j+ K(1+iC_1)(Z-\hat W_j),
\end{align}
where each cluster variable $\hat W_j,\, j=1\dots n$ is equal to all variables in the $j$-th cluster and the mean field is given by 
\begin{equation}\label{eq:mf}
   Z=\frac{1}{N}\sum_{j=1}^n N_j\hat W_j.
\end{equation}

The rotational symmetry is a continuous symmetry and can be removed by introducing a corotating frame. Indeed, passing to polar coordinates $\hat W_j =R_j\text{e}^{i\phi_{j}}$ we can use phase differences $\phi_{1}-\phi_j$, $j=2\dots n$ as new variables. In this way we obtain a {\it reduced system} of $n$ real amplitudes and $n-1$ phase differences that does not have a phase shift symmetry any more. Note that in this system periodic solutions of the form $\vec W(t) =\vec W_0 e^{i\Omega t}$, which due to phase shift invariance appear generically in the original system \eqref{eq:GCSLO}, turn into equilibria. Similarly, there are quasiperiodic solutions of the original system \eqref{eq:GCSLO} that turn into periodic solutions (limit cycles) of the reduced system. 

\paragraph{Transversal and longitudinal instabilities, cluster splitting, and continuum limit}
A fundamental observation of cluster dynamics in systems of coupled identical oscillators is, that sizes and number of the observed clusters depends on the choice of the parameters.  
The reduced cluster system we introduced above can only be used to find $n$-cluster solutions with predefined fixed cluster sizes $N_1\dots N_n$. This limitation can be overcome, by considering the cluster sizes as additional free parameters. For large systems it is natural to consider the relative cluster sizes $N_j/N$ as real parameters in the interval $[0,1]$. This will allow us to use them together with the other system parameters in a numerical bifurcation analysis. Note that in this way the overall system size $N$ has disappeared from the system, but along the resulting branches of solutions with varying cluster size only rational points can be realized in a finite system of corresponding commensurable size. 
In this context also a vanishing relative cluster size makes sense. It corresponds to a solitary oscillator, which in the limit of large $N$ has a vanishing contribution to the mean field \cite{Schulen.2022}.

Note that for $n>1$ such a system also includes $n-1$-cluster solutions that appear after merging two clusters of size $N_j$ and $N_\ell$ into a single one of size $N_j+N_\ell$. In this way, the reduced cluster system has again invariant subspaces that are characterized by equal amplitudes and phase differences for the corresponding merged clusters. Thus, a bifurcation analysis of a given reduced cluster system may reveal not only bifurcations where the cluster structure remains unchanged, but also bifurcations where such a merging occurs. All  bifurcations within a given reduced cluster system  are often called {\em longitudinal}, since the corresponding critical subspaces of the bifurcation are longitudinal to the given cluster subspace.

Describing instead a bifurcation that induces a further splitting of a given cluster structure is much more complicated, because the resulting cluster structure is not known a priori. Indeed, there can be bifurcations, where in a cluster splitting multiple branches of $n+1$-cluster solutions bifurcate simultaneously. Also bifurcations where a cluster decays immediately into more than two subclusters or even into solitary oscillators have been observed \cite{PhysRevE.99.062201}.
Whenever from a branch of solutions in an $n$-cluster subspace a secondary branch bifurcates that is not contained in this subspace, such a bifurcation is referred to as a  {\em transversal} bifurcation, since its critical subspace is transversal to the given cluster subspace.

A good method to detect also transversal instabilities in a bifurcation analysis of a reduced cluster system, without going back to the full system, is to include a single additional oscillator as a so called {\em test oscillator} that is driven by the mean field \eqref{eq:mf} without contributing to it. To probe the transversal stability of cluster $j$ we use the corresponding reduced cluster system with the additional equation
\begin{equation}
\label{eq:test1}
\partial_t  V_j = V_j - (1+iC_2) \big| V_j \big|^2 V_j+ K(1+iC_1) (Z- V_j).
\end{equation}
The resulting system has an invariant subspace with $ V_j=\hat W_j$, where the dynamics coincide with that in the original cluster subspace. But a linear stability analysis can detect instabilities transversal to this subspace (see \cite{Ku.2015, Pikovsky.2001b}). 

In the following sections we will use the approaches described above to give a detailed picture of the emergence of stable two and three cluster solutions in system \eqref{eq:GCSLO}. For 2-cluster solutions we use the relative cluster size parameter 
$\rho_1:=N_1/N\in[0,1]$. For the secondary cluster splitting, i.e., the transition from 2-cluster solutions with sizes $N_1,N_2$ to 3-cluster solutions with sizes $N^\prime_1, N^\prime_2, N^\prime_3$, we assume that $N^\prime_1=N_1$ and $N_2=N^\prime_2+ N^\prime_3$ such that it is convenient to use the splitting ratio $$\rho_2:=N^\prime_2/(N^\prime_2+N^\prime_3)\in[0,1].$$

We will now perform a comprehensive numerical bifurcation analysis of the reduced 2-cluster and 3-cluster systems and investigate the respective transversal stabilities/instabilities by employing corresponding test oscillators. 

\paragraph{Numerical methods\label{subsec:numerics}}
We perform extensive numerical simulations of the complete oscillator system to complement our analytical approach. We use Python's Scipy package to numerically integrate the system. We use the 'Zvode' integrator with the 'Adams' method and a time step $dt= 0.01$ \cite{Virtanen.2020}. After a transient period of 3000 steps, the system is further integrated for 1000 steps to ensure that we capture the long-term dynamics. We use the 'Scipy.cluster.hierarchy' package to study the cluster distribution of the solution and order the clusters by their maximal amplitude. For numerical bifurcation analysis, we use the continuation software AUTO-07p \cite{Doedel.1981} and 
the Julia package NLsolve.jl.
\paragraph{Abbreviations\label{subsec:abb}} A list of abbreviations used in the text is given in the appendix \ref{Abbreviations}.

\section{\label{subsec:ReducedC} A Complete Picture of stable 2-Cluster solutions}
\begin{figure}
\includegraphics{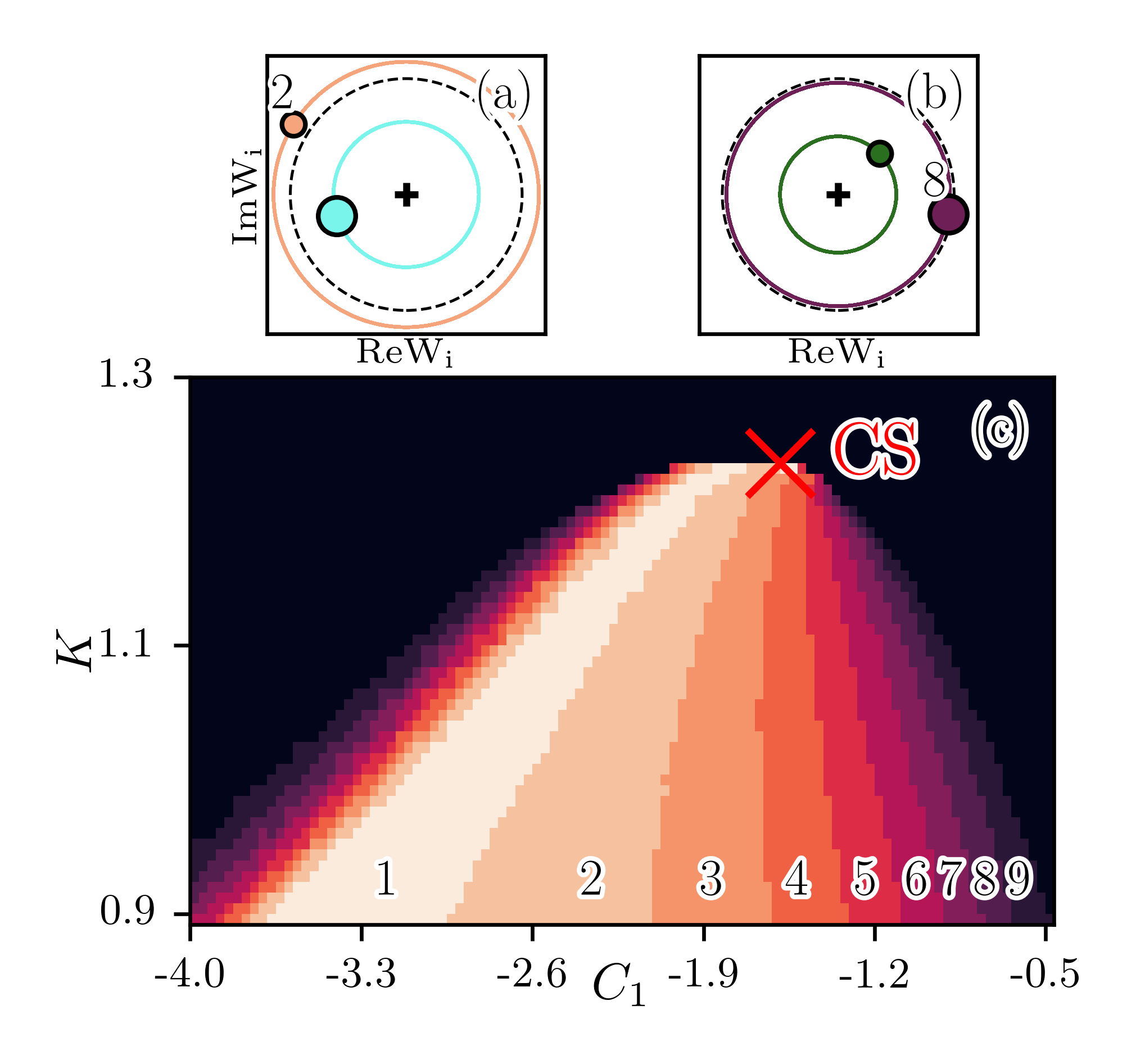}
\caption{\label{fig:Heatmap10} Simulations of system \eqref{eq:GCSLO} with N=10. Panels \textbf{(a)} and \textbf{(b)} show snapshots of cluster solutions in the complex plane with $N_1=2$ and  $N_1=8$, respectively (Parameters $C_1$ and $K$ from the corresponding regions in panel \textbf{(c)}.
Panel \textbf{(c)}: Typical size $N_1$ of the high amplitude cluster for stable 2-cluster solutions with varying parameters $K$ and $C_1$ close to the cluster singularity (CS). The black region indicates the stable synchronous state.}
\end{figure}
With the discovery of the cluster singularity, the question of how simple 2-cluster equilibra are organized has been solved. To understand how these solutions bifurcate to more complex cluster configurations and more complex temporal behavior as the coupling strength $K$ is decreased and the completely incoherent zero-mean state is approached, we study their dynamics in the reduced 2-cluster system. We demonstrate how the cluster singularity manifests itself in this reduced space and then investigate the longitudinal instabilities of the 2-cluster solutions. 

First, we use numerical simulations with 10 oscillators to visualize the fan-like arrangement of stable 2-cluster solutions, which is a fingerprint of the cluster singularity.
Therefore, we integrated Eqs. (\ref{eq:GCSLO}) in the parameter region indicated by the dashed square in Figure \ref{fig:PrevRe}. For each parameter value, we used 500 different initial conditions. For every initial condition, we analyzed the cluster distribution. We determined the average number of oscillators in the high-amplitude cluster for a given parameter tuple and rounded the average value to an integer. We used a resolution of 0.035 along the $K$-axis and 0.008 along the $C_1$-axis. The result is shown in Figure \ref{fig:Heatmap10}. 
Panels \textbf{(a)} and \textbf{(b)} show two different snapshots of 2-cluster solutions that both have 2 oscillators in one cluster and 8 oscillators in the other cluster. To discriminate between the two cases, we label the clusters $N_1:N_2$ and follow the convention that $N_1$ indicates the number of oscillators in the cluster with the higher maximal amplitude. Hence, (a) shows a $2:8$ and (b) an $8:2$ cluster solution. In Figure \ref{fig:Heatmap10} (c), we show where the different 2-cluster solutions are dominant in the parameter plane. In the black region synchronous oscillations prevail. 
The numbers in colored regions indicate the cluster size $N_1$ of the most probable 2-cluster state and reveal their organization in the parameter plane. For values of $K$ below the cluster singularity (CS), there is a region of stable 2-cluster solutions where with increasing $C_1$ the size of the high amplitude cluster increases stepwise.

\subsection{\label{subsec:Reduced2D}Organization of 2-cluster solutions at the cluster singularity}
\begin{figure}
\includegraphics{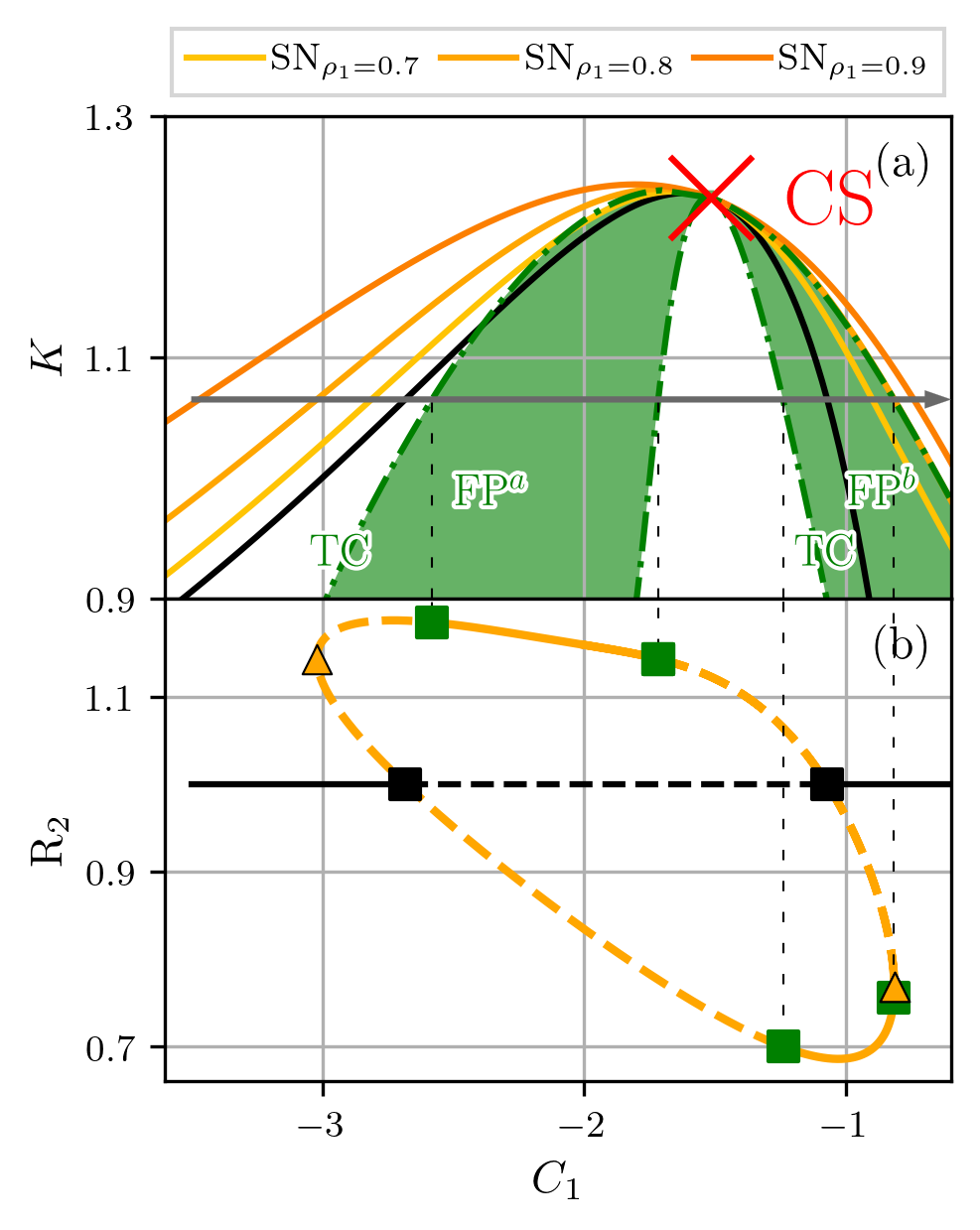}
\caption{\label{fig:TypeICS} Bifurcation diagrams of fixed point solutions of the reduced 2-cluster system \eqref{eq:2cl_1}--\eqref{eq:2cl_3} for $C_2=2$ close to the cluster singularity (CS, red cross). \textbf{(a)}: Two-parameter bifurcation diagram. Black line: Benjamin-Feir instability; orange lines: saddle-node bifurcations for different relative cluster sizes $\rho_1$; green dash-dotted lines: transverse, transcritical bifurcations (TC) stabilizing and destabilizing the 2-cluster solutions for $\rho_1=0.8$ (see text). Green areas: regions in which one of the two 2-cluster fixed points generated in the SN bifurcations is stable for $\rho_1=0.8$. Notice that the rightmost TC bifurcation line does not lie on the right branch of the $\text{SN }{\rho_1=0.8}$ but just close to it. 
 \textbf{(b)}: One-parameter bifurcation diagram for $K=1.065$. Orange lines: 2-cluster fixed point solutions; black line: synchronous fixed point solution. Solid lines indicate stable and dashed lines unstable solutions. $\blacktriangle$: SN bifurcations;  $\blacksquare$: TC bifurcations.}
\end{figure}

The reduced system for 2-cluster solutions reads\cite{GarciaMorales.2012}:
\begin{align}
\partial_t R_1 =& R_1 (1 - K - R_1^2 )-C_1 K (\rho_1 - 1) R_2\sin\phi_{12} \label{eq:2cl_1}\\ 
 &+ K \rho_1 (R_1 - R_2 \cos\phi_{12}) + K R_2 \cos\phi_{12},\nonumber \\
\partial_t R_2 =& R_2 - R_2^3 
-K \rho_1 \left(C_1 R_1 \sin\phi_{12} - R_1 \cos\phi_{12} + R_2\right)  \label{eq:2cl_2}\\
\partial_t \phi_{12} =& -\frac{1}{R_1 R_2}
\left [ C_1 K \left(\rho_1 \left( (R_1^2 + R_2^2) \cos\phi_{12} - 2 R_1 R_2 \right) \right.  \right. \label{eq:2cl_3}\\
&\left. + R_2 (R_1 - R_2 \cos\phi_{12}) \right )+ K R_2^2 \sin\phi_{12}\nonumber \\
&\left.  -(R_1 - R_2) (R_1 + R_2) (C_2 R_1 R_2 + K \rho_1 \sin\phi_{12}\right)].\nonumber
\end{align}
The solutions in Fig. \ref{fig:Heatmap10}\textbf{(a), (b)} are fixed points of Eqs. \eqref{eq:2cl_1}--\eqref{eq:2cl_3}. 
The synchronous fixed point, characterized by $R_1=R_2=1\text{ and } \phi_{12}=0$ exists for all parameter values. However, it is stable only for $K>-2(1+C_1C_2)/(1+C_1^2)$, i.e. above the Benjamin-Feir (BF) curve\cite{Nakagawa.1994}, as illustrated in Figure \ref{fig:PrevRe} for $C_2=2$. Additionally, a pair of fixed points (FP), $(R_1^a,R_2^a,\phi_{12}^a)$ and  $(R_1^b,R_2^b,\phi_{12}^b)$ exists for each distinct value of $\rho_1$  and $C_2$ in a certain region of the parameter plane $K-C_1$\cite{Kemeth.2019}.
This pair of FPs appears and vanishes in saddle-node (SN) bifurcations. 
This can be seen in Figure \ref{fig:TypeICS}, where one and two-parameter bifurcation diagrams for selected $\rho_1$ values are shown, similar to the approach in \cite{Kemeth.2020}. 

Distinct SN-bifurcations for three different values of $\rho_1$, which create pairs of 2-cluster solutions, are shown  in different orange tones in Figure \ref{fig:TypeICS}(a). Notice the umbrella-like fanning out of the different SNs: the more unbalanced the two clusters are, the further the SN lies inside the region where the synchronized solution is stable. This is consistent with the numerical results of Figure \ref{fig:Heatmap10}. The different SNs touch the BF curve, shown as the black solid line, tangentially at the cluster singularity, depicted by the red cross. Recall that there is a continuum of SN bifurcations as $\rho_1$ is continuously increased from 0 to 1. 

Studies of the original system \eqref{eq:GCSLO} showed that at the SN all  2-cluster solutions, except for the most unbalanced solution with one cluster containing a single, solitary oscillator, are transversely unstable \cite{Kemeth.2019,Kemeth.2020}. Since the transverse stability cannot be seen in the reduced system, we assessed the stability of the 2-cluster fixed points using a test oscillator as described above. In this way, we could determine the transcritical bifurcations (TC, green dash-dotted curves) in which 2-cluster fixed points lose their stability, but only unstable 3-clusters emerge\cite{Kemeth.2019,Kemeth.2020}. 
In Figure \ref{fig:TypeICS}, parameter regions in which the 2-cluster solutions with relative size $\rho_1=0.8$ is stable are shown in green. These regions roughly overlap with the $N_1=2$ and $N_1=8$ tongues in Figure $\ref{fig:Heatmap10}$, but do not match precisely because the stability regions of the different 2-cluster solutions overlap, while in Figure $\ref{fig:Heatmap10}$ we plotted the averaged cluster size obtained from 500 simulations. 

In the one-parameter bifurcation diagram Figure \ref{fig:TypeICS} \textbf{(b)} we plotted a branch of 2-cluster fixed point solutions  with $\rho_1=0.8$, showing $R_2$ for varying parameter $C_1$ and fixed $K=1.065$ (orange curve). The black line shows the synchronous solution. Dashed lines indicate unstable solutions, whereas solid lines indicate stable solutions. The 2-cluster fixed points are born in SN bifurcations (indicated by orange triangles). Both branches interact with the synchronous solution in the Benjamin-Feir instability (indicated by the black square), however at different parameter values.
Along the branches of 2-cluster fixed point solutions with $\rho_1=0.8$ in Figure \ref{fig:TypeICS} \textbf{(b)} we find both solutions where $N_1$ refers to the high-amplitude cluster and to the low-amplitude cluster. They exchange their role exactly at the Benjamin-Feir instability. The labels '$a$' and '$b$' for the stability regions in the upper panel refer to this property, i.e., the cluster of size $N_1$ having the higher or lower amplitude, respectively (cf. Fig.~\ref{fig:Heatmap10}(a) and (b)).



\subsection{\label{subsec:Stab2Cl}Further bifurcations within the 2-cluster subspace}
Having outlined the stability of the stationary 2-cluster solutions close to the cluster singularity, we discuss now their longitudinal stability for further decreasing $K$. Previous studies\cite{Kemeth.2018, Kemeth.2019, Ku.2015} suggest that they may lose stability through longitudinal Hopf bifurcations (HB), which we analyze in detail below. By keeping $C_2=2$ and $\rho_1=0.8$, we can add additional bifurcations to the bifurcation diagram shown in Figure \ref{fig:TypeICS}.
\begin{figure*}
\includegraphics{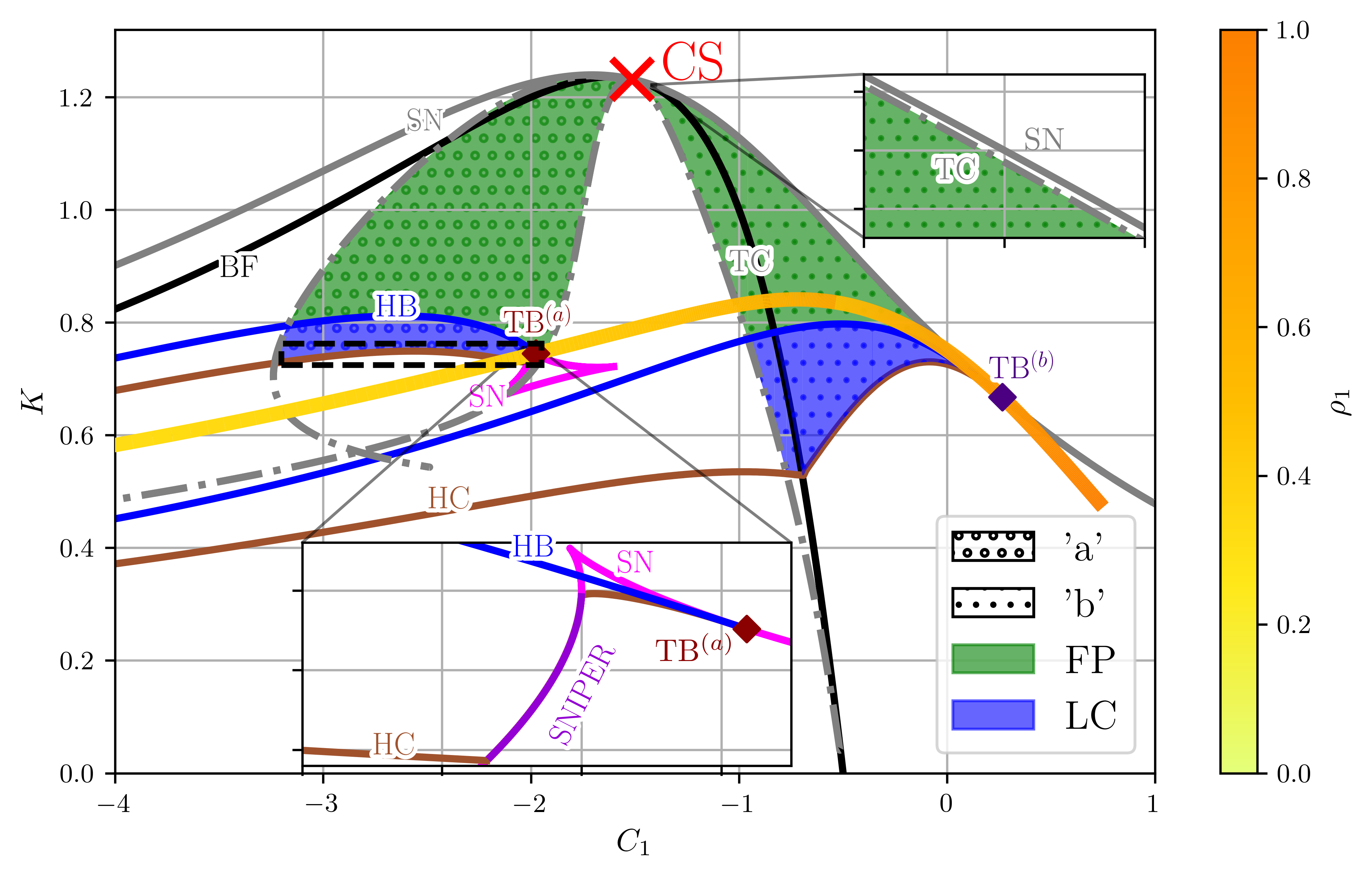}
\caption{\label{fig:Review2Cl} Bifurcation diagram with $C_2=2$ and $\rho_1=0.8$ depicting how 2-cluster LC solutions are created and destroyed in the $K-C_1-\text{plane}$. The black line represents the Benjamin-Feir (BF) instability and indicates where the synchronous solution loses stability. The SNs and TCs are shown in gray solid and dash-dotted lines, respectively. The top right inset certifies that the TC lies under the SN. The green areas show where the 2-cluster fixed points are stable. Filled circles show where the cluster distribution 'a' is stable and hollow points where the cluster distribution 'b' is stable. The blue lines represent longitudinal Hopf bifurcations (HB) creating LC solutions, whereas the brown lines indicate homoclinic bifurcations (HC) destroying these LC solutions. The 2-cluster LCs are stable in the blue areas. The pink lines represent 3 SN bifurcations creating new pairs of fixed point solutions. SNs, HCs, and HBs meet in a Takens-Bogdanov (TB) point. The bottom inset shows a magnification of the bifurcation diagram close to the TB of fixed point 'a'. The orange-yellow color graded line $\rho_1$ changes from 1 to 0 and shows the location of the Takens-Bogdanov points. The parameter region inside the dashed rectangle will be treated in Figure \ref{fig:HC_CS}.}
\end{figure*}
The top part of Fig.~\ref{fig:Review2Cl} for $K>0.9$ is a simplified representation of Figure \ref{fig:TypeICS}. We concentrate on the two parameter regions where fixed points 'a' and 'b' are transversely stable. As $K$ decreases, both stable fixed point solutions undergo a supercritical longitudinal Hopf bifurcation, indicated by the blue lines, giving rise to stable LC 'a' respectively 'b' (in the corotating frame). At even lower values of $K$, these LCs are destroyed in homoclinic (HC) bifurcations, depicted by the brown lines. The blue areas indicate where the LCs are longitudinally stable within the reduced 2-cluster system. Next, we will focus on the HC of LC 'b'. In a HC, a limit cycle collides with a saddle point. On the right side of the BF bifurcation, LC 'b' collides with the unstable fixed point 'a'. On the left side of the BF bifurcation, the saddle that destroys the LC solution is the unstable synchronous solution. The change of the saddle point involved in the HC bifurcation,  a so called {\em Bykov T-point} \cite{Bykov.1993, Knobloch.2014}, leads to the kink in the HC line at $C_1 \approx -0.7$. SN, HB, and HC tangentially meet in a Takens-Bogdanov (TB) point, depicted by the indigo diamond. 

Turning now to LC 'a', we observe again that the HC and HB meet in a TB point. However, the bifurcation fine structure around this point is more intricate then fo LC 'b'. The SN associated with the TB, shown in pink, is not the one that creates the 2-cluster fixed point solutions discussed above, but one that generates a further pair of unstable fixed points. It forms a closed contour in the $K-C_1$ parameter plane involving three cusp points. For our discussion, these novel fixed points are of no further importance and will not be explored further. Instead, we look at how the existence region of the LC 'a' closes towards large values of $C_1$. As can be seen in the inset of Figure \ref{fig:Review2Cl}, the HC bifurcation destroying the LC 'a' first interacts with the pink SN in a Saddle-Node Loop that renders the SN a SNIPER (Saddle-Node of infinite period). In a second Saddle-Node Loop, the SNIPER turns back into a simple SN, and a second HC is created. This HC then ends in a TB point, marked by a brown diamond, where it interacts with the HB and the SN. 

A comparison of the bifurcation scenario of LCs 'a' and 'b' reveals that their existence regions lie in both cases between a HB and a HC that emerge from a TB bifurcation. In this sense, we can say that the TB organizes both the longitudinal instability of the fixed point solution and the existence region of the LCs. Until now, we merely investigated the two different types of solutions associated with $\rho_1=0.8$; however, for a system of many oscillators, a multiplicity of relative cluster sizes exist. To extend these observations to different $\rho_1$-values, we continue the codimension-two TB-point in three parameters $C_1,\,K$ and $\rho_1$. 
The TB-point is characterized as a fixed point where the Jacobian has a zero eigenvalue with an algebraic multiplicity two. We implemented this via the Routh-Hurwitz criterion for the Hopf bifurcations \cite{Liu.1994}. 
In Figure \ref{fig:Review2Cl}, the orange to yellow color scale represents the $\rho_1$-values varying along the TB branch. For $\rho_1=1$, the solitary state case, the TB-point can be found analytically because one radial variable decouples from the system of equations, see Appendix \ref{app:solitary_TB}. The existence of a TB point for 2-cluster solutions of all possible size ratios $\rho_1$ ensures that all stable 2-cluster fixed point solutions emerging at the cluster singularity undergo a longitudinal Hopf bifurcation to a stable 2-cluster LC that is in turn destroyed at lower values of $K$ by a HC. However, so far, the stability of the 2-cluster LCs is not guaranteed in the full model. We can only be sure that they are stable in the immediate neighborhood of the Hopf bifurcation. The further fate of the LC solution is examined in the next section, where we show that they transversally branch into 3-cluster solutions in an intricate way.

\section{\label{sec:3ClSM}3-Cluster Solutions}
In this section, we increase the dimension of our reduced system and admit 3-cluster solutions. 
The corresponding system of five equations for the amplitudes $R_1,\,R_2,\,R_3$ and the phase differences $\phi_{12},\,\phi_{13}$
is given in the Appendix \ref{app:3ClMan}. We now use this set of equations to investigate both the transverse stability of the above-discussed 2-cluster LC solutions and the longitudinal stability of the resulting 3-cluster solutions. 

\subsection{\label{sec:rho2_02}Splitting of 2-Cluster Periodic Solutions}
\begin{figure*}
    \centering
    \includegraphics{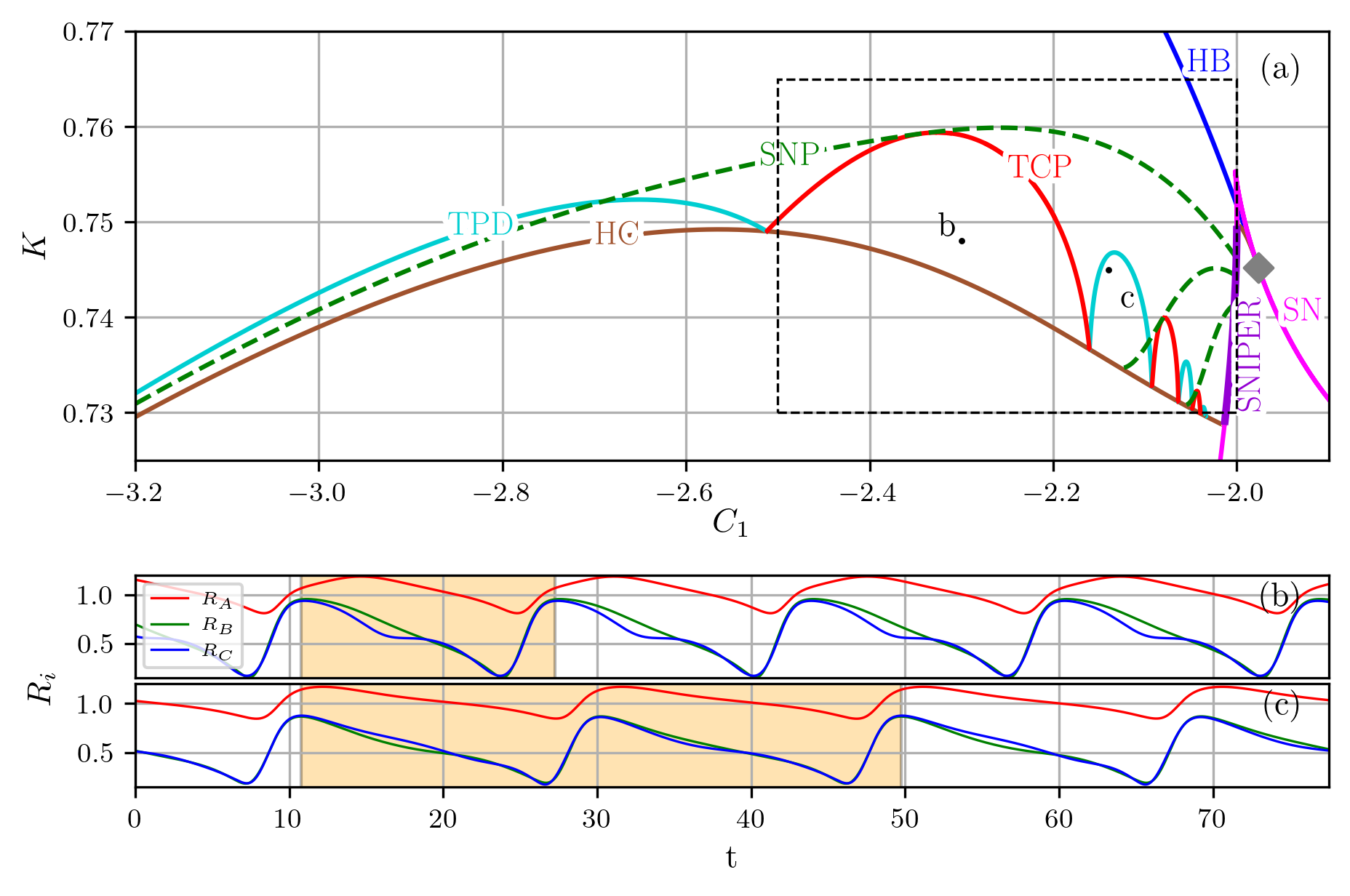}
    \caption{\textbf{(a)} Bifurcation diagram of Eqs.\eqref{eq:2cl_1}--\eqref{eq:2cl_3} with $C_2=2 \text{, } \rho_1=0.2 \text{ and } \rho_2=1$ displaying how 2-cluster LC solutions lose transverse stability. Solid lines indicate bifurcations that involve 2-cluster solutions, whereas dashed lines show bifurcations in which only 3-cluster solutions are involved. The blue, brown, and pink lines show the Hopf, homoclinic, and SN bifurcations of 2-cluster solutions, and the gray diamond symbol shows the location of the TB point. All these bifurcations are also shown in Figure \ref{fig:Review2Cl}. Red lines give the locations of transverse, transcritical bifurcations of the 2-cluster LC solutions and the cyan ones, those of transverse period doubling (TPD) bifurcations. The green lines indicate saddle-node bifurcations of periodic 3-cluster orbits (SNP). \textbf{(b)} Time series of the amplitude variables $R_i$ of 3-cluster LC solution at $K=0.748, C_1=-2.3$, $\rho_2=1$ (indicated by a point labeled 'b' in (a). \textbf{(c)} Time series of the amplitude variables $R_i$ of 3-cluster LC solution at $K=0.745, C_1=-2.14$, $\rho_2=0.6$ (indicated by a point labeled 'c' in (a). $\rho_2=0.6$ has been chosen for visibility reasons. For both time series, one oscillation period is highlighted by an orange background color. The parameter region inside the dashed rectangle will be treated in Figure \ref{fig:FullDimensionalSystem}.}
    \label{fig:HC_CS}
\end{figure*}

First, we focus on the transverse stability of the small amplitude cluster of the 2-cluster LC-solution 'a'. For $\rho_2=1$, the reduced 3-cluster system describes the case of a solitary oscillator in the third cluster, which in the case of a 2-cluster solution amounts to the situation of a test oscillator. Figure \ref{fig:HC_CS} depicts the bifurcation scenario for parameter values in the dashed box \tikz[baseline=-0.5ex]{\draw[dashed] (0,0) -- (0.5,0);}, ($C_1 \in [-3.2,-2.1]$)  of Figure \ref{fig:Review2Cl}. In Figure \ref{fig:HC_CS}, solid lines indicate bifurcations that involve 2-cluster solutions, whereas bifurcations in which only 3-cluster solutions participate are shown as dashed lines.
The lines denoted by HB, HC, and SNIPER indicate bifurcations of the 2-cluster LC solutions and correspond to those also shown in Figure \ref{fig:Review2Cl}. In the entire $C_1$ interval, the 2-cluster LCs that are born in the HB become transversely unstable before they are destroyed in the HC. The break-up of the 2-cluster LC occurs in one of two different symmetry breaking bifurcations: in a transcritical bifurcation of periodic orbits (TCP) or in a transverse period-doubling (TPD) bifurcation. In the former ones, shown as red lines in Figure \ref{fig:HC_CS}, the 2-cluster LC interacts with an unstable 3-cluster LC, which renders one cluster of the 2-cluster LC solution unstable with respect to transverse perturbations.   
The unstable 3-cluster LC participating in the transcritical bifurcation is born in a saddle-node of periodic orbits bifurcations (SNP), shown in green in Figure \ref{fig:HC_CS}. 

The cyan lines mark the locations of TPDs of the 2-cluster LC. In these TPDs, the low-amplitude cluster of the 2-cluster LCs  lose their transverse stability, and a stable 3-cluster LC branch that oscillates with twice the period is born. 
Time traces of the amplitudes of stable 3-cluster LCs involved in the two scenarios are shown in Figure \ref{fig:HC_CS} \textbf{(b)} and \textbf{(c)}, respectively. In both cases, 
the trajectories of clusters 'B' and 'C' (blue and green), which resulted from the splitting   are still close, while the high-amplitude cluster 'A' has kept its integrity and its trajectory is practically unaffected. 
In \textbf{(c)}, the splitting comes together with a period-doubling
where the two clusters 'B' and 'C' show an opposite behavior in each cycle, while again, the high-amplitude cluster 'A' has hardly changed its dynamics. 
Below, we refer to the cluster 'A', which is not affected by the transverse instability but keeps its integrity, as the "observer cluster". 

Both the transverse transcritical and the transverse period-doubling bifurcations interact with the homoclinic bifurcation in a codimension-2 point. There are several of such codimension-2 points along the curve of homoclinic bifurcations. Note that for $C_1\approx>-2.02$, there are further TCP and TPD bifurcation lines that have not been resolved in Figure \ref{fig:HC_CS}. We assume that this alternation of TPD, TCP, and codimension-2 points could be organized by some intricate type of symmetry-breaking homoclinic bifurcation. 

In addition, the lines of TCPs touch the SNP lines tangentially, which is reminiscent of the behavior at the above-discussed cluster singularity, but with 2- and 3-cluster LCs involved instead of 1- and 2-cluster fixed points: disregarding the observer cluster, the transcritical bifurcation in which one cluster becomes transversely unstable is reminiscent of the BF instability, in which the (1-cluster) synchronous solution becomes transversely unstable. Moreover, the 3-cluster LC solutions that interact with the TCP are born in SNPs, which reminds of the 2-cluster fixed point solutions that interact with the BF, born in SNs. In the following section, we will demonstrate that the point of contact between SNP and TCP indeed forms a new type of cluster singularity that organizes 3-cluster LC solutions, similar to the cluster singularity that organizes 2-cluster fixed point solutions. 

\subsection{\label{subsec:TIICS} Type-II Cluster Singularity}
\begin{figure}
    \centering
    \includegraphics{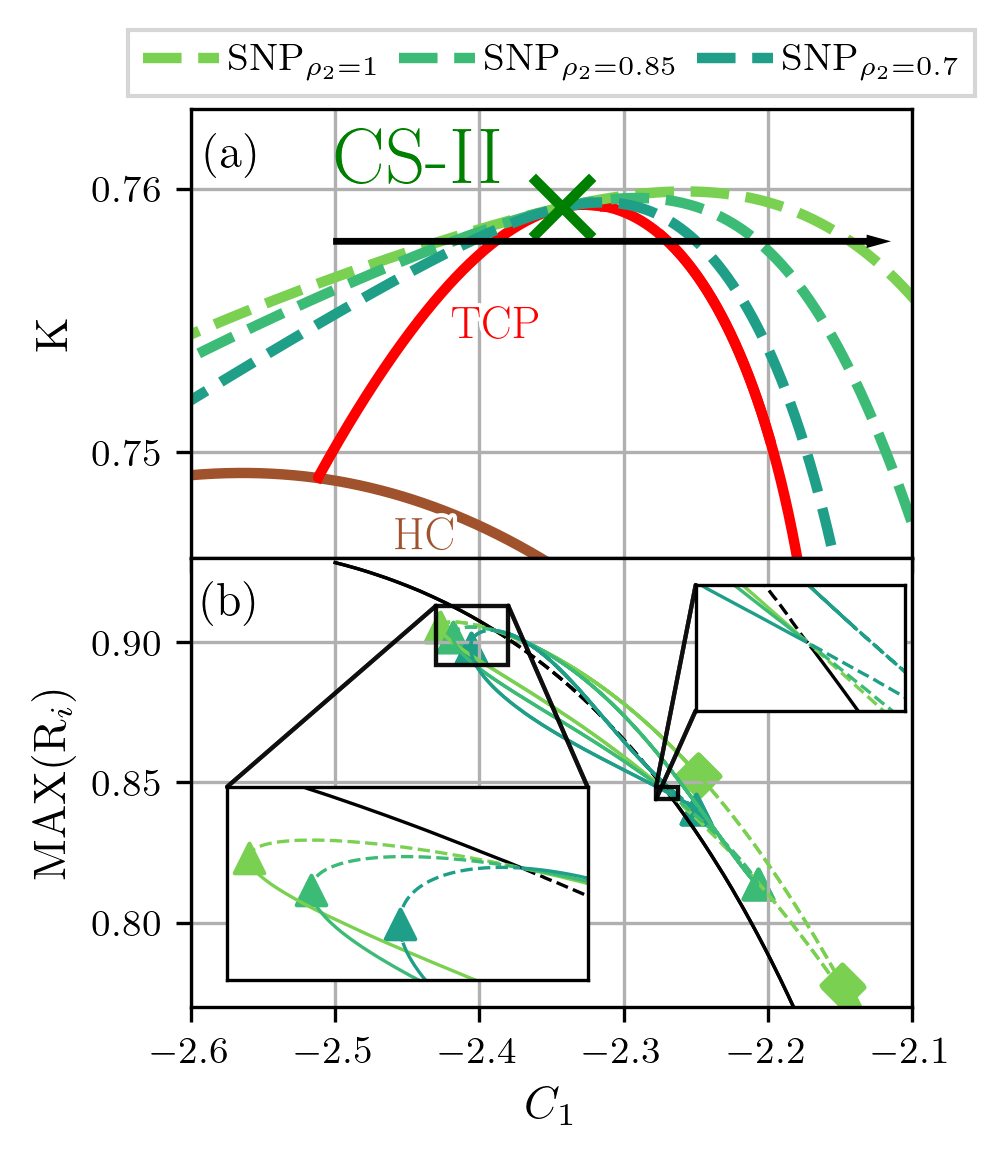}
    \caption{\textbf{(a)} Bifurcation diagram of 2-cluster and 3-cluster LCs in the $K-C_1$ parameter plane close to the Type-II cluster singularity with $C_2=2$, $\rho_1=0.2$ and different values of $\rho_2$. As in Figure \ref{fig:HC_CS}, dashed lines indicate bifurcations of 3-cluster solutions and solid lines those of 2-cluster solutions; green lines show locations of SNPs for different values of $\rho_2$, red and brown lines show the locations of transverse transcritical and homoclinic bifurcations of 2-cluster solutions, 
    respectively.
    The SNPs and the TCP touch tangentially in a point, that we dub \textbf{Type-II cluster singularity} CS-II. 
    The black arrow indicates where the 1-parameter bifurcation diagram shown in (b) is located ($K=0.758$). \textbf{(b)} One-parameter bifurcation diagram showing $R_2$ respectively $R_3$ for the 2- and 3-cluster solutions as a function of $C_1$. Dashed lines show unstable solutions and filled lines show stable solutions; 3-cluster limit-cycle solutions are shown in green and 2-cluster limit-cycle solutions in black. 
    $\blacktriangle$: SNPs, \rotatebox{45}{$\blacksquare$} TCPs or longitudinal PD bifurcations of 3-cluster solutions. Note that at the lower $C_1$ values SNPs and 3-cluster transcritical bifurcations (not shown as \rotatebox{45}{$\blacksquare$} are so close that they seem to fall together).}
    \label{fig:TypeIICS}
\end{figure}

To obtain further insight into the arrangement of 3-cluster solutions in the parameter plane, we calculated the location of two further SNPs creating 3-cluster solutions for different values of $\rho_2$, i.e., different relative size of the second and third cluster, keeping $\rho_1$, i.e., relative size of the observer cluster, constant. They are shown as dashed green curves in Figure \ref{fig:TypeIICS} \textbf{(a)} in a region of the $C_1-K$ plane, which is a magnification of the region around the largest 'TCP arc' in Figure \ref{fig:HC_CS} \textbf{(a)}, 
but without showing the TPDs for the sake of clarity.
The different SNPs all touch the TCP tangentially and apparently at the same point. Below, we demonstrate that they indeed interact in a codimension-2 singularity that organizes the formation of all different 3-cluster limit-cycle solutions for a given relative size $\rho_1$ of the observer cluster. 
We coin this novel codimension-2 point a \textbf{Type-II cluster singularity} because of its similarity to the cluster singularity\cite{Kemeth.2018}, introduced by Kemeth et al., which organizes the formation of all 2-cluster fixed point solutions. From now on we will refer to the latter as Type-I cluster singularity. 
\\
We further illustrate the similarities between the Type-I and the Type-II cluster singularities with the one-parameter bifurcation diagram along $C_1$ at a constant value of $K=0.758$ depicted in Figure \ref{fig:TypeIICS} \textbf{(b)}. The maxima of the radius $R_2$ of the 2-cluster solutions are shown in black, those of $R_3$ of the 3-cluster solutions for different values of $\rho_2$ in different green shades. The 2-cluster solution is stable for low and high values of $C_1$ (black solid line), but the second cluster is transversely unstable at intermediate values (dashed black line). Pairs of 3-cluster solutions are born in SNPs (triangular symbols); each of the two 3-cluster LCs interacts with the transcritical bifurcation of the 2-cluster solution and is stabilized in a transverse, transcritical bifurcation (diamond symbols) in some $C_1$ interval, except the 3-cluster solitary solution which is born stable. Notice that the SNP and the stabilizing TCP are very close in parameter space and cannot be distinguished in part \textbf{(b)} of Figure \ref{fig:TypeIICS}. The observer cluster stays transversely stable for all parameters shown and thus does not split into two parts in any of the discussed transverse bifurcations. Having this in mind and comparing Figures \ref{fig:TypeIICS} and \ref{fig:TypeICS} the similarity between the two scenarios is apparent.\\

Yet, the existence of the observer cluster also implies an essential difference between the two types of cluster singularities: While the Type-I cluster singularity is of codimension two in the system parameters $C_1,\,C_2,\,K$ and provides a unique point in the 
 $K-C_1$ parameter plane, the 
 Type-II cluster singularity depends also on the relative size $\rho_1$ of the observer cluster. This means it is of codimension two in the parameters $C_1,\,C_2,\,K,\,\rho_1$, such that for fixed $C_2$  it appears as a curve in the $K-C_1-\rho_1$ parameter space.

\subsubsection*{Characterization of the Type-II cluster instability by the transversal Floquet spectrum} 
So far, we have described the Type-II cluster singularity a point of tangential intersection of different SNP curves. Now we will describe it as a singular point along each such curve. Recall that a characteristic property of a cluster singularity is that induces a cluster splitting where the new clusters resulting from the splitting are stable (cf. Fig.~\ref{fig:Heatmap10}). To assess this transversal stability, we have to employ the reduced 3-cluster system with an additional test oscillator, which leads to a 7-dimensional system.
With this system, we repeat the two-parameter continuation of the SNPs in the $K-C_1$ plane (cf. Figure \ref{fig:TypeIICS}) for different values of $\rho_2$ and monitor the Floquet multipliers. Since we are continuing along SNP solutions, we expect two Floquet multipliers to be equal to 1. The first arises from the time-shift invariance of the LC, representing the neutral direction along the solution. The second is associated with the saddle-node bifurcation\cite{Guckenheimer.1983}. For a generic point on a SNP curve there are, in addition to these two neutrally stable multipliers, four more stable ones. The last one indicates the transversal stability of the solution. This multiplier is characterized by the fact that the corresponding Floquet mode is transversal to the 3-cluster subspace, i.e., it has a component that separates the test oscillator from the cluster to which it is attached.
\begin{figure}
    \centering
    \includegraphics{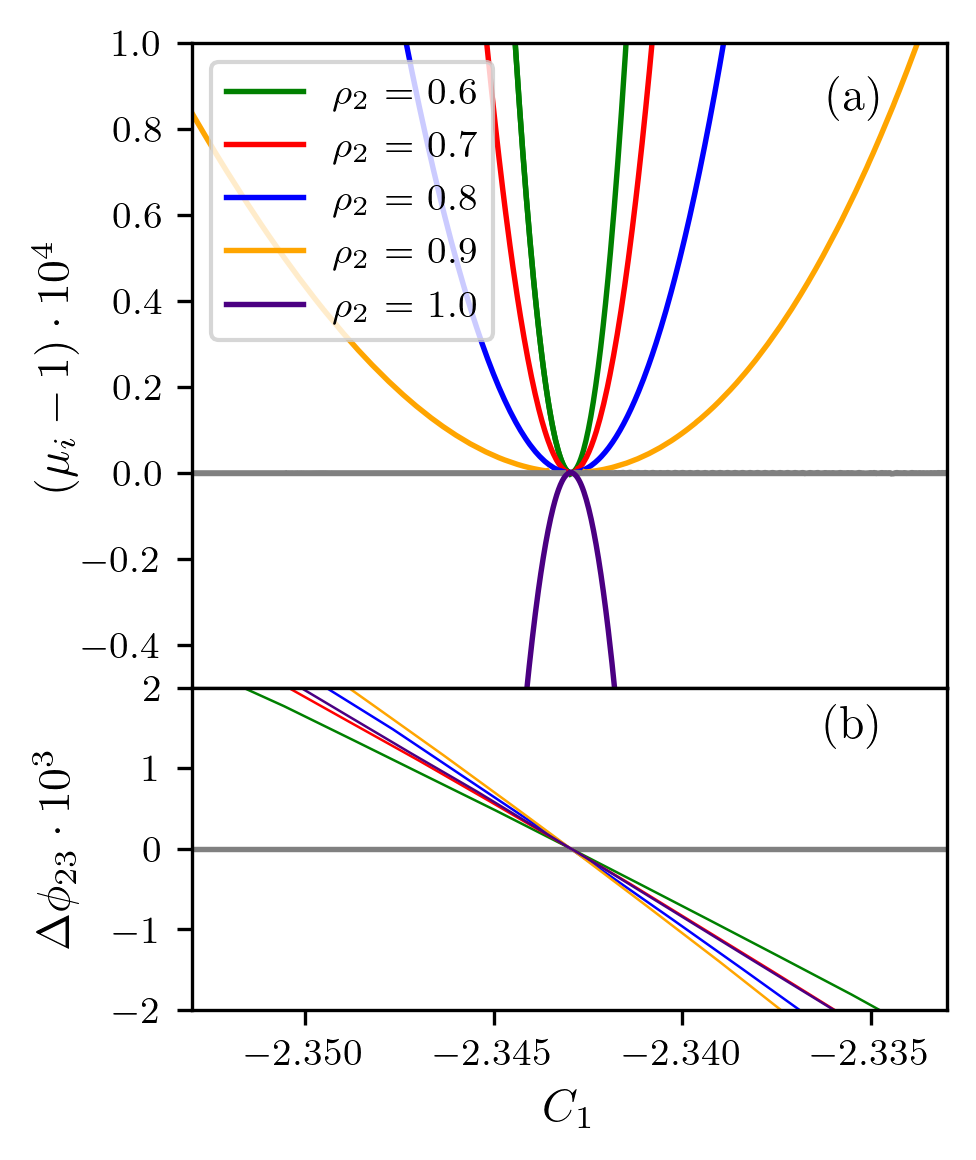}
    \caption{\textbf{(a)} Projection of the trivial and transverse Floquet Multiplier $\mu_i$ of the LC solutions along SNP bifurcations in the $K-C_1$ plane onto the $C_1$ direction for different values of $\rho_2$ with $C_2=2$ and $\rho_1=0.2$. The Floquet mulitpliers are obtained using the Auto-07p.
    \textbf{(b)} Phase difference between the second and third clusters along the SNPs for different values of $\rho_2$.}
    \label{fig:FMs}
\end{figure}
The Floquet multipliers of the transversal and the neutrally stable directions are plotted in Figure \ref{fig:FMs} versus $C_1$. Note that in this continuation $K$ is adapted to $C_1$ in order to select in the $K$-$C_1$ plane a specific SNP curve with a fixed value of $\rho_2$. Close to the singular point all transverse Floquet multipliers are all unstable, except for the limiting case 
$\rho_2 = 1$. This case corresponds to a solitary oscillator in one of the new clusters, such that no further splitting instability is possible. All other curves of multipliers show a quadratic tangency to the criticality at $\mu=1$.
A similar behavior has also been found by analyzing the center manifold of the Type-I cluster singularity \cite{Kemeth.2020}.

To give another representation of the degeneracy of the bifurcation point, in Figure \ref{fig:FMs}\textbf{(b)}, we have plotted $\Delta \phi_{23}$ along the SNPs. It can be seen that for all values of $\rho_2$, $\Delta \phi_{23}$ vanishes at the same point. This point coincides with the location of the cluster singularity in \textbf{(a)}. Hence, at the cluster singularity, the second and third clusters merge, which underlines that this point corresponds to a 2-cluster LC on the TCP curve that has an additional instability. The Floquet multiplier analysis thus confirms that the Type-II cluster singularity is a codimension-2 bifurcation that is characterized by the following properties:
\begin{itemize}
    \item All different 3-cluster solutions corresponding to a particular relative size $\rho_1$ of the observer cluster merge to a specific degenerate 2-cluster solution with relative cluster size $\rho_1$.
    \item In the corresponding reduced 3-cluster system with an additional test oscillator, there are 3 Floquet multipliers equal to 1 simultaneously. They correspond to a saddle-node/transcritical bifurcation that co-occurs with an additional transverse instability. Both of these bifurcations are associated with a Floquet multiplier 1. The third Floquet multiplier equal to 1 results from the phase shift symmetry of the underlying periodic solution. 
    \item The cluster with the relative size $\rho_1$ keeps its integrity since it is neither affected by the saddle-node bifurcation nor by the transcritical bifurcation. It thus takes on the role of an 'observer'.
    \item There is a multitude of Type-II cluster singularities, being on the one hand associated with different values of $\rho_1$, but also for a particular value of $\rho_1$ more than one Type-II cluster singularity might exist (cf. Figure \ref{fig:HC_CS}\textbf{(a)}). 
\end{itemize}
This behavior has analogies with the behavior at the Type-I cluster singularity \cite{Kemeth.2019}, with the following differences: (a) for the Type-I cluster singularity the observer cluster does not exist; (b) instead of LC solutions FPs are involved; (c) there is only one Type-I cluster singularity in the $C_1-K$ plane. 

\subsection{\label{sec:overview}Overall picture of 2- and 3-cluster solutions}
\begin{figure*}
    \centering
    \includegraphics{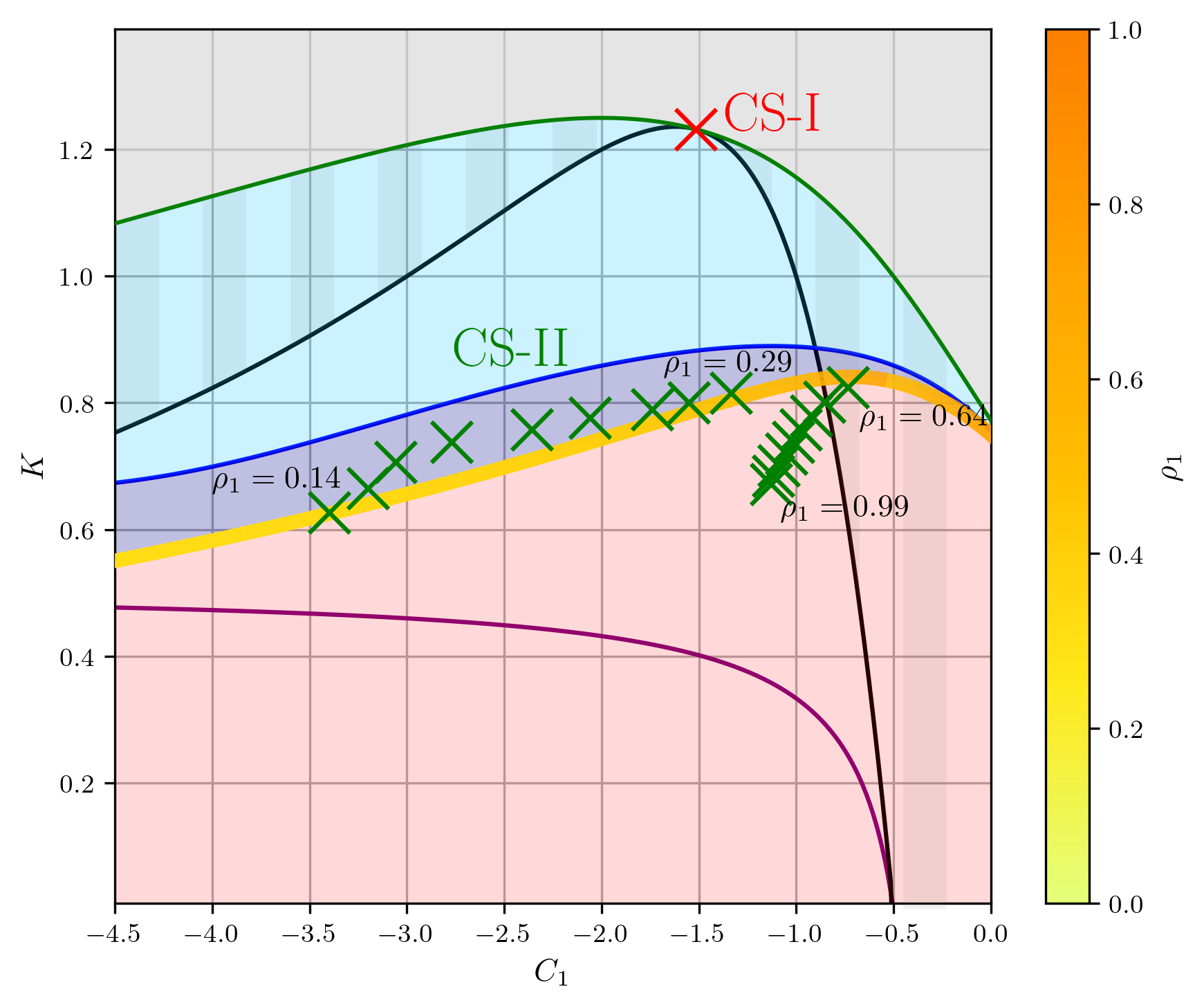}
    \caption{\label{fig:Conclusion}Bifurcation diagram of a globally coupled Stuart-Landau oscillators system under linear global coupling for $C_2=2$. The black line depicts the Benjamin-Feir (BF) instability, indicating that the synchronous solution loses stability. The synchronous solution is globally stable in the gray area and coexists with cluster solutions in the gray-striped, colored regions. The purple-colored line indicates the stability boundary of the vanishing-mean solution. The green line indicates the saddle-node bifurcation creating the solitary 2-cluster solution. The blue line represents an estimation for the longitudinal HB giving rise to stable 2-cluster LC solutions in the dark blue region. The orange-colored line indicates a series of Takens-Bogdanov points that organize the emergence of LC solutions. The color bar indicates the respective $\rho_1$ value. The red cross indicates the presence of the Type-I cluster singularity. The green crosses indicate the position of the Type-II cluster singularities ranging from $\rho_1 \in [0.14;0.29]$ and $\rho_1 \in [0.64;0.99]$. The red region indicates the presence of more complex solutions, such as higher-order cluster solutions and chimera states.}
\end{figure*}

With the results on 2- and 3-cluster solutions discussed above, we can extend the stability diagram of Figure \ref{fig:PrevRe}. This is done in Figure \ref{fig:Conclusion}, where we added several regions with known dynamics.
The green line marks the SN of solitary 2-cluster FP solutions, where stable 2-cluster solutions appear for the first time. The point where this SN touches the BF bifurcation, shown in black, is the Type-I cluster singularity. The fan-like organization of the other 2-cluster solution guarantees, on the one hand, that above this line, there are no other 2-cluster solutions. On the other hand, when changing $C_1$ for a fixed $K$-value from a value on the SN to the right of the Type-I cluster singularity to a value on the left of it, we come along stable 2-cluster solutions of all possible relative sizes. We colored in light blue the region where stable 2-cluster FP solutions for some relative size exist. Note that this region has a tremendous multistability of 2-cluster FPs, which is not resolved in this overall picture.
In the gray-hatched regions, the synchronous solution coexists with other solutions. E.g. in the hatched light-blue regions the synchronous solution coexists with the FP 2-cluster solutions.

When decreasing $K$, all these 2-cluster FP solutions undergo a longitudinal supercritical Hopf bifurcation, implying that there is some boundary below which oscillatory 2-cluster solutions exist. We estimated the location of this boundary by continuing the HBs of several different 2-cluster solutions and approximated their envelope (shown as a dark blue line). Thus, within the dark blue region, 2-cluster LCs exist but possibly coexist, particularly close to the border, with 2-cluster FP attractors. 

The orange colored line marks the position of Takens-Bogdanov bifurcations involving 2-cluster fixed point solutions, parametrized by the relative cluster size  $\rho_1$. Thus, for each value of $\rho_1$, there is not only a HB creating 2-cluster LCs, but also a HC destroying them again. However, the HCs are irrelevant for stable solutions since all of the 2-cluster LCs become unstable before in a transverse bifurcation in which 3-cluster LCs are involved. 
The region in parameter space where stable 3-cluster solutions appear can be approximated by the positions of the Type-II cluster singularities for different values of $\rho_1$. 
As argued above, at the Type-II cluster singularity, three Floquet multipliers simultaneously become equal to 1. This suggests that, in principle, Type-II cluster singularities could be continued in the $K-C_1-\rho_1$-parameter space. However, standard continuation software does not provide this option. Therefore, we determined the position of the type-II cluster singularity for discrete values of $\rho_1$ instead. They are shown as green crosses in Fig. \ref{fig:Conclusion}. We see that they split into two branches, one where the crosses are above the TB line and the other one where they are below that line. Along the first branch, $\rho_1$ was varied between $0.14$ (at approx. $C_1=-3.45$) and $0.29$ ($C_1\approx-1.4$), in the second group between $0.99$ and $0.64$, suggesting that there are two $\rho_1$ intervals, $\rho_1 \in [0.14;0.29]$ and $\rho_1 \in [0.64;0.99]$, for which a continuum of Type-II cluster singularities exists. For $\rho_1$ values outside these intervals, we have numerical evidence that the cluster singularities do not exist. Hence, roughly, stable 3-cluster LCs exist close to the envelope of the type-II cluster singularities. Their existence region is considerably smaller than the one of 2-cluster solutions, and they, in part, coexist with more complex dynamics, as chaotic 3-cluster solutions that emerge in longitudinal PD cascades or 'higher' cluster solutions with more than 3 clusters.
For $K$-values beyond the three cluster region and above the border to completely incoherent behavior, the dynamics become exceedingly complex and have not been explored further in this study.

\section{\label{sec:FDS}Typical Behaviour in the Original System}
To validate our approach using the reduced manifold, we compared it with data from the original system \eqref{eq:GCSLO}. We set N = 20 and select a region in the $K-C1$ parameter plane where where in the reduced system 2 cluster solutions with $\rho_1=0.2$ were stable, which thus corresponds to a 2-cluster solution with four oscillators in one cluster for $N=20$. As previously mentioned, the main difficulty with working with the full system is the critical coexistence of dynamic solutions. To circumvent this problem, we use again a statistical approach. For a given set of $K-C_1$ pairs, we integrate the system 100 times with different random initial conditions and determine the cluster distribution. We furthermore order the clusters according to their amplitude and can then do a statistical evaluation of the obtained cluster structures. This approach does not directly describe the attractor with the largest basin of attraction, but provides a reasonable estimate.

\begin{figure*}
    \centering
    \includegraphics{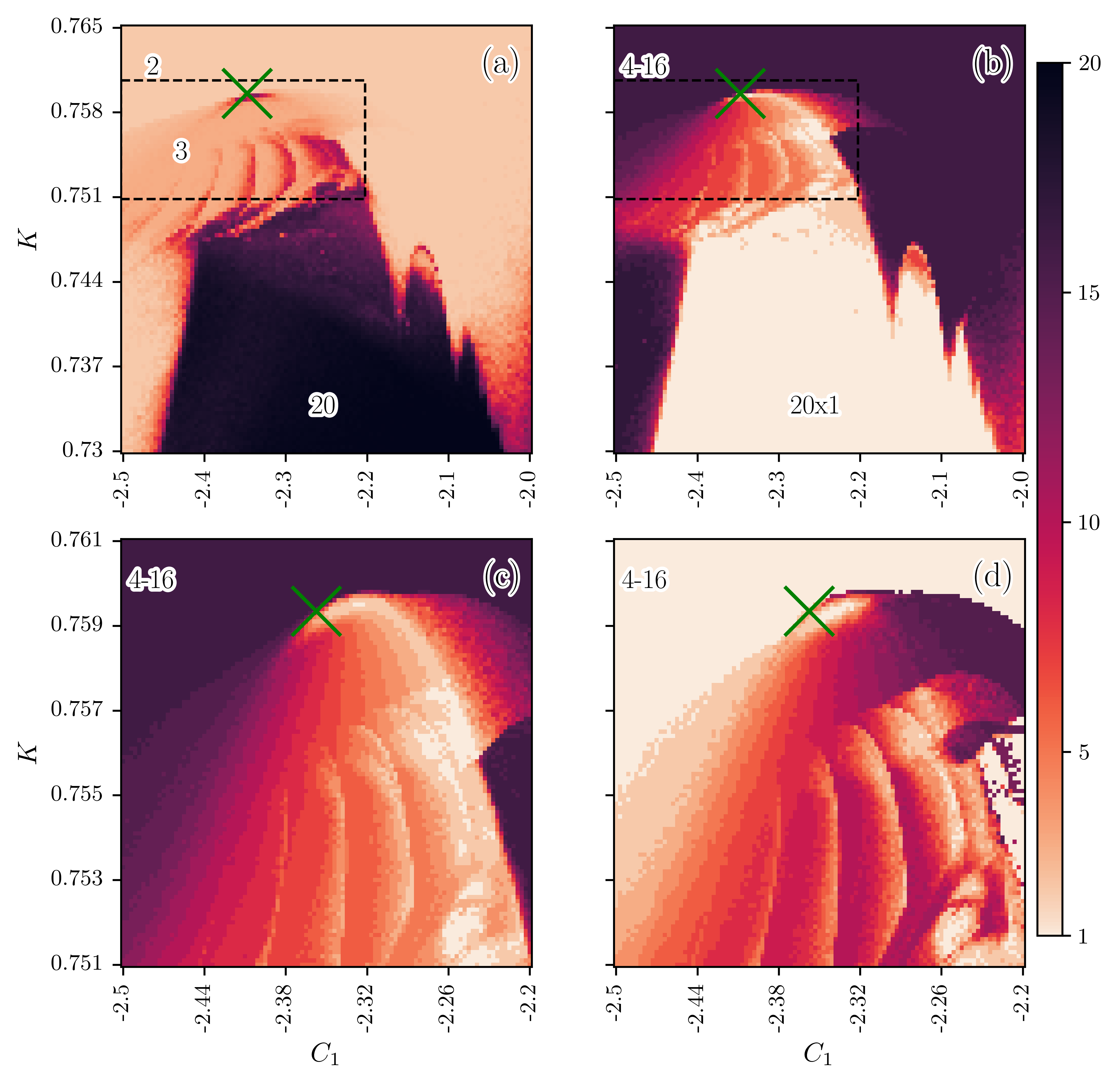}
    \caption{For every pixel, corresponding to one pair of $(K-C_1)$ parameters, the full dynamical system of Eqs. (\ref{eq:GCSLO}) has been integrated for 4000-time steps and 100 different initial conditions and $N=20$. The average cluster solution has then been plotted. The average number of clusters is plotted in \textbf{(a)}. The number of oscillators in the cluster with the second highest maximal amplitude has been plotted in \textbf{(b)}. The white region corresponds to an incoherent solution, and the dark purple plots the 16 oscillators in the 4-16 cluster solution. \textbf{(c)} is a zoomed version of the dashed box in \textbf{(b)} and \textbf{(d)} is identical to \textbf{(c)}, but the number of oscillators in the cluster with the third highest maximum amplitude is plotted.}
    \label{fig:FullDimensionalSystem}
\end{figure*}

In Figure \ref{fig:FullDimensionalSystem} we show the results of the simulations in a parameter region around the Type-II cluster singularity (green cross) for $\rho_1=0.2$ (see the rectangle  indicated in Figure \ref{fig:HC_CS}). Consider first plate \textbf{(a)} where the average number of clusters is plotted.
At the top and the right border, there is a region where the 2-cluster solutions prevail. The border of this region towards lower values of $K$ reminds of the hill- or tongue-like curves of the transcritical and period-doubling bifurcations in Figure \ref{fig:HC_CS}, and reflects that the 2-cluster solutions are destabilized through those bifurcations. Below the largest hill-like structure at the top left, 3-cluster solutions dominate the dynamics; for lower $K$ values, higher cluster solutions break up until complete incoherence, depicted by the black region.

More information about the different cluster solutions can be obtained from plate \textbf{(b)}, where the average number of oscillators in the cluster with the second highest amplitude is shown. In the following, we refer to it in short as the second cluster. The purple color in the 2-cluster region stands for 16 oscillators, which is in accordance with our expectation for $\rho_1=0.2$. For $N=20$, this corresponds to a 4-16 cluster. In the hill-like structure where 3-cluster solutions exist, the fanning out of the colors is reminiscent of the behavior seen in the neighborhood of the Type-I cluster singularity in Figure \ref{fig:Heatmap10}. A magnification of this fan is displayed in plate \textbf{(c)}. In plate \textbf{(d)}, the average number of oscillators in the cluster with the third highest amplitude is shown. We see that in an intermediate stripe of this region, the number of oscillators in the second cluster indeed decreases from 15 to 1 when decreasing $C_1$, while the one in the third cluster increases from 1 to 15. This is a manifestation of the Type-II cluster cluster singularity that is located close to the top of the hill, adding up to 16 for each parameter value within the fan, while there are always 4 oscillators in the observer cluster. Hence, when increasing $C_1$ from the left border of the 3-cluster region where the 4-15-1 cluster exists, we transverse a series of solutions with one oscillator less in the second cluster and one more in the third one until we reach the 4-1-15 solution. 
Interestingly, Figure \ref{fig:FullDimensionalSystem} \textbf{(a)}-\textbf{(d)} reveal some more structure that we did not observe in the reduced system: From many 3-cluster region, a new tongue seems to emerge. At this point, we can only speculate that this might be the manifestation of even another type of cluster singularity, that organizes 4-cluster solutions. 

\section{\label{sec:OaC}Conclusion and Outlook}
In this work, we focused on cluster formation in a system of Stuart-Landau oscillators under linear mean-field coupling. We built upon the previously introduced cluster singularity using a reduced manifold approach. We showed that a continuous line of Takens-Bogdanov points in the $K-C_1-\rho_1$ parameter space organizes the emergence of 2-cluster LC solutions. These LC solutions are the prerequisites for stable 3-cluster LC solutions, while stable 3-cluster FP solutions do not seem to exist. The 3-cluster LC solutions emerge from a transverse period-doubling bifurcation of the 2-cluster LCs or are organized by a codimension-2 point, lying on a transcritical bifurcation in which the 2-cluster LC becomes unstable. We coined this point Type-II cluster singularity. In both cases, the high-amplitude cluster, which we call the observer cluster, stays intact, and the low-amplitude cluster splits up. A Type-II cluster singularity may exist for different sizes of the observer cluster and generate all possible 3-cluster LCs that may arise through the break-up of the low-amplitude cluster. The integrity of the observer cluster therefore naturally introduces a hierarchical structure to the cluster structure.

While we have been able to uncover basic mechanisms of how stable 3-cluster solutions emerge in globally coupled SL oscillators, many other questions have arisen that we have not addressed here. An obvious question is why the Type-II cluster singularities exist in two distinct intervals of the relative cluster size $\rho_1$, or, in other words, what distinguishes those values from the ones where no cluster singularity is observed (cf. Figure \ref{fig:Conclusion}). Probably related is the question of why the Type-II CS in one interval lies above the line of TB points, in the other one below it. 
In this context, it would be very helpful if the numerical obstacle we faced when trying to continue the Type-II cluster singularities in the $ K-C_1-\rho_1$ parameter space could be solved. The challenge here lies in continuing the LC solutions. Another open problem is what causes or organizes the occurrence of alternations between transverse period-doubling and transcritical bifurcations through which the 2-cluster LCs are destabilized, and how many Type-II cluster singularities can exist for a given $\rho_1$ (cf. Figure \ref{fig:HC_CS})? \\
Furthermore, our direct numerical simulations provide evidence of the emergence of higher cluster solutions ($n>3$), which emerge in tongues from the different 3-cluster solutions in one 'fan' (\ref{fig:FullDimensionalSystem}). It appears, therefore, worthwhile to investigate whether a cascade of cluster singularities exists that organizes the emergence of all cluster solutions hierarchically. It will, however, be hardly possible to resolve the cascade numerically since we expect the parameter intervals for which they exist to scale geometrically with the number of clusters, similar to the scaling of successive period-doubling bifurcations in the Feigenbaum cascade, and other techniques are needed to approach the problem. A further obstacle to the investigation of this question is that the 3-cluster LCs are in part destabilized by longitudinal period doubling or Neimark-Sacker bifurcations. These processes create dynamically complex solutions that are challenging to track and make the bifurcation analysis even more challenging. \\

From the symmetry perspective, the existence and unfolding of the two types of cluster singularities, as well as their hierarchical organization, are a consequence of the equivariance of Eqs. (\ref{eq:GCSLO}) under the symmetry group $\mathbf{S}_N$. This means that similar bifurcation diagrams, particularly cluster singularities, should exist in any system of globally coupled 2-dimensional oscillators, not just globally coupled Stuart-Landau oscillators considered here. Furthermore, using the tool of equivariance under the full permutation group $\mathbf{S}_N$, as, e.g., done for polynomial vector fields of at most cubic order in \cite{Fiedler.2021, Stewart.2003, Golubitsky.2002, Barabash.2021, Dias.2003}, will further help understand the intricate dynamics of globally coupled, identical oscillators.


\section*{\label{sec:Ack}Acknowledgement}
The authors thank Alexander Gerdes for fruitful discussions.

\section*{\label{sec:Ref}References}
\bibliography{Emergence3Cl}
\appendix

\section{\label{app:solitary_TB} Determining the Takens-Bogdanov Point for the solitary cluster}
In a Takens Bogdanov bifurcation, a saddle-node bifurcation, a Hopf bifurcation and an homoclinic bifurcation all meet in one point. 
When $\rho_1=1$, Eq. \eqref{eq:2cl_1} decouples from \eqref{eq:2cl_2} and \eqref{eq:2cl_3} and reads:
\begin{align*}
\partial_t R_1 &= R_1 (1-R_1^2).
\end{align*}
This equation has two positive fixed points, R$_1=1$ and R$_1=0$. We discard the second FP and substitute R$_1=1$ into the other two solutions, resulting in:
\begin{equation}
\label{app:eq:2Deq}
\begin{cases}
\begin{aligned}
\partial_t R_2 &= -C_1 K \sin (\phi_{12} )-R_2 \left(K+R_2^2-1\right)+K \cos (\phi_{12} ) \\
\partial_t \phi_{12} &=\frac{C_1 K (R_2-\cos (\phi_{12} ))+C_2 R_2 \left(R_2^2-1\right)-K \sin (\phi_{12} )}{R_2},
\end{aligned}
\end{cases}
\end{equation}
where we can find the FP solutions using the Mathematica command $\texttt{Solve}$. There are six fixed point solutions in total, where three are unphysical with negative R$_2$ values; one corresponds to the synchronous solution, and the last two solutions correspond to the two 2-cluster solutions. One of these fixed points \textbf{X}$_1 = (R_2^*, \phi_{12}^*)$ is given by:  
\begin{equation}
\label{app:eq:FP}
\begin{cases}
\begin{aligned}
R_2^* =& \frac{\sqrt{-\mathcal{A}-4 C_1 K-2 K+5}}{\sqrt{10}}\\
\phi_{12}^*=& \arctan\left(\frac{5+\mathcal{A}-(K+2C_1(5+\mathcal{A}-4K-KC_1))}{(-2+C_1)(5+\mathcal{A}+2K+4KC_1)}\right)
\end{aligned}
\end{cases}
\end{equation}
where $\mathcal{A} =\sqrt{-4 (C_1-2)^2 K^2-20 (2 C_1+1) K+25}$.
To calculate the saddle-node curve, we linearize Eqs. (\ref{app:eq:2Deq}) around \textbf{X}$_1$ and set the determinant of the jacobian to zero, implying
\begin{equation}
\label{app:eq:Det}
2 (\mathcal{A}-10) (2 C_1+1) K+5 (\mathcal{A}+5)-4 (C_1-2)^2 K^2=0
\end{equation} 
The Hopf curve can be obtained similarly by setting the trace of the Jacobian to zero. The resulting condition for the Hopf bifurcation reads:
\begin{equation}
\label{app:eq:Tr}
\mathcal{A}+(4 C_1-3) K=0.
\end{equation} 
The Takens Bogdanov point can hence be obtained by solving the system of equations where both (\ref{app:eq:Det}) and (\ref{app:eq:Tr}) are equal to 0. These two lines intersect in the $K-C_1$ plane at the point with coordinates: $C_1^{TB}\rightarrow0.75$ and $K^{TB }\rightarrow2\sqrt{5}-4$. This is in accordance with Figure \ref{fig:Review2Cl}.
\section{\label{app:3ClMan}Reduced 3-Cluster Subspace}
Applying the transformation to polar coordinates with the phase reduction results in the following system of equations describing the dynamics in the 3-cluster subspace. 
\begin{widetext}
\begin{equation}
\begin{cases}
\begin{aligned}
\partial_t R_1 &= C_1 K (\rho_1 - 1) \left((\rho_2 - 1) R_3 \sin(\phi_{13}) - \rho_2 R_2 \sin(\phi_{12})\right) - R_1 (K + R_1^2 - 1) \\
&\quad + K \rho_1 (R_1 - R_3 \cos(\phi_{13})) + K (\rho_1 - 1) \rho_2 (R_3 \cos(\phi_{13}) - R_2 \cos(\phi_{12})) + K R_3 \cos(\phi_{13}), \\
\partial_t R_2 &= -C_1 K \rho_1 R_1 \sin(\phi_{12}) + K (\rho_1 - 1) (\rho_2 - 1) R_3 \sin(\phi_{13}) (C_1 \cos(\phi_{12}) + \sin(\phi_{12})) \\
&\quad + K (\rho_1 - 1) (\rho_2 - 1) R_3 \cos(\phi_{13}) (\cos(\phi_{12}) - C_1 \sin(\phi_{12})) + K \rho_1 R_1 \cos(\phi_{12}) \\
&\quad + K \rho_2 R_2 - K \rho_1 \rho_2 R_2 - K R_2 - R_2^3 + R_2, \\
\partial_t R_3 &= -K \rho_1 \left(C_1 R_1 \sin(\phi_{13}) - R_1 \cos(\phi_{13}) + R_3\right) \\
&\quad + K (\rho_1 - 1) \rho_2 \left(-C_1 R_2 \sin(\phi_{12} - \phi_{13}) - R_2 \cos(\phi_{12} - \phi_{13}) + R_3\right) - R_3^3 + R_3, \\
\partial_t \phi_{12} &= -\frac{C_1 K \rho_1 R_1^2 \cos(\phi_{12}) - C_1 K \rho_1 R_1 R_2 + C_1 K \rho_2 R_1 R_2 - C_1 K \rho_1 \rho_2 R_1 R_2}{R_1 R_2} \\
&\quad + \frac{K (\rho_1 - 1) (\rho_2 - 1) R_3 \cos(\phi_{13}) (C_1 (R_1 \cos(\phi_{12}) - R_2) + R_1 \sin(\phi_{12}))}{R_1 R_2} \\
&\quad + \frac{K (\rho_1 - 1) (\rho_2 - 1) R_3 \sin(\phi_{13}) (C_1 R_1 \sin(\phi_{12}) - R_1 \cos(\phi_{12}) + R_2)}{R_1 R_2} \\
&\quad - \frac{C_1 K \rho_2 R_2^2 \cos(\phi_{12}) + C_1 K \rho_1 \rho_2 R_2^2 \cos(\phi_{12}) + C_2 R_1^3 R_2 - C_2 R_1 R_2^3 + K \rho_1 R_1^2 \sin(\phi_{12})}{R_1 R_2} \\
&\quad + \frac{K \rho_2 R_2^2 \sin(\phi_{12}) - K \rho_1 \rho_2 R_2^2 \sin(\phi_{12})}{R_1 R_2}, \\
\partial_t \phi_{13} &= \frac{C_1 K \left(\rho_1 (2 R_1 R_3 - (R_1^2 + R_3^2) \cos(\phi_{13})) - (\rho_1 - 1) \rho_2 (-R_1 R_2 \cos(\phi_{12} - \phi_{13}) + R_3 (R_1 - R_3 \cos(\phi_{13})) + R_2 R_3 \cos(\phi_{12}))\right)}{R_1 R_3} \\
&\quad + \frac{R_3 (R_3 \cos(\phi_{13}) - R_1) - (R_1 - R_3) (R_1 + R_3) (C_2 R_1 R_3 + K \rho_1 \sin(\phi_{13}))}{R_1 R_3} \\
&\quad + \frac{K (\rho_1 - 1) \rho_2 \left(-R_1 R_2 \sin(\phi_{12} - \phi_{13}) + R_2 R_3 \sin(\phi_{12}) + R_3^2 (-\sin(\phi_{13}))\right) - K R_3^2 \sin(\phi_{13})}{R_1 R_3}.
\end{aligned}
\end{cases}
\end{equation}
\end{widetext}
\section{List of Abbreviations \label{Abbreviations}}

\renewcommand{\arraystretch}{1.2} 
\setlength{\tabcolsep}{10pt}     

\begin{minipage}{0.5\textwidth} 
\begin{tabular}{@{}p{0.2\textwidth}p{0.7\textwidth}@{}}
\toprule
\textbf{SL} & Stuart-Landau \\
\textbf{FP} & Fixed point \\
\textbf{LC} & Limit cycle \\
\textbf{BF} & Benjamin-Feir \\
\textbf{SN}  & Saddle-node \\
\textbf{TC} & Transcritical \\
\textbf{HB} & Hopf bifurcation \\
\textbf{TPD} & Transverse period doubling \\
\textbf{SNIPER} & Saddle-node of infinite period \\
\textbf{SNP} & Saddle-node of periodic orbits \\
\textbf{TCP}  & Transcritical bifurcation of periodic orbits \\

\bottomrule
\end{tabular}
\end{minipage}
\end{document}